\documentclass[superscriptaddress,preprint,showpacs,preprintnumbers,amsmath,amssymb]{revtex4}
\usepackage[dvips]{graphicx}
\usepackage{amsmath}
\usepackage{bm}

\begin{document}

\preprint{BA-03-14}

\title{Flipping $SU(5)$ Towards Five Dimensional Unification}

\author{I. Dorsner}
\email{idorsner@ictp.trieste.it}
\affiliation{
The Abdus Salam International Centre for Theoretical Physics\\
Strada Costiera 11, I-34100 Trieste, Italy} \affiliation{
Bartol Research Institute\\
University of Delaware\\
Newark, DE 19716, USA}

\begin{abstract}
It is shown that embedding of flipped $SU(5)$ in a
five-dimensional $SO(10)$ enables exact unification of the gauge
coupling constants. The demand for the unification uniquely
determines both the compactification scale and the cutoff scale.
These are found to be $M_C \approx 5.5 \times 10^{14}\,{\rm GeV}$
and $M_* \approx 1.0 \times 10^{17}\,{\rm GeV}$ respectively. The
theory explains the absence of $d=5$ proton-decay operators
through the implementation of the missing partner mechanism. On
the other hand, the presence of $d=6$ proton-decay operators
points towards the bulk localization of the first and the second
family of matter fields.
\end{abstract}

\pacs{11.10.Hi,12.10.Dm,12.10.Kt}

\maketitle

\newpage
\section{Introduction}
The main motivation for supersymmetry (SUSY), besides its ability
to stabilize the Higgs mass against the radiative corrections, is
the way it steers the gauge couplings, within the Minimal
Supersymmetric Standard Model (MSSM), towards the unification at a
very high energy scale ($M_{\rm GUT}$). Assuming this is not an
accident but a signal for a new physics we are prompted not only
to embrace the MSSM but to incorporate it into the grand unified
theory (GUT) where the gauge unification represents a genuine {\em
prediction\/} of the framework. Another genuine {\em prediction\/}
of the true GUT is, of course, a proton decay. It turns out,
however, that it is very problematic to build both realistic and
simple SUSY GUT scheme and still preserve the exact gauge coupling
unification. For example, the parameter space of the simplest of
all such schemes, the minimal $SU(5)$ SUSY GUT, has been severely
constrained by the experimental limits on proton lifetime
\cite{Goto:1998qg,Babu:1998ep,Murayama:2001ur,Bajc:2002bv,Bajc:2002pg}.

The crux of the problem is that the exact gauge unification requires
threshold corrections. But to create these corrections one needs certain
fields, responsible for the proton decay, to be too light compared to the
existing experimental constraints unless an {\em ad hoc\/}
tuning of parameters takes place \cite{Bajc:2002bv,Bajc:2002pg}. This
problem was not so serious in the
past since the low-energy values of the gauge couplings were not known well
enough, leaving a lot of room for maneuvering. The situation has
changed after the electroweak precision measurements
and the improvements in measurements of the strong coupling constant. The
error bars have simply become sufficiently small to prevent the exact
unification without the help of the troublesome threshold corrections.
So, the question of whether we can achieve the exact gauge unification in
accord with the low-energy measurements in a natural manner within
SUSY GUTs is something we have to address.

Among the fields that can improve on the gauge unification, via
threshold corrections, are the familiar colored Higgsinos. These
are the fields that are responsible for $d=5$ proton-decay
operator. Therefore, one wants them light enough to generate the
appropriate corrections but heavy enough to avoid violation of the
experimental limits on proton lifetime. This, again, is an
extremely difficult task. One can entirely avoid the need of
satisfying these conflicting requirements by using a flipped
$SU(5)$ group~\cite{DeRujula:1980qc,Georgi:1980pw,Barr:1981qv}
which automatically explains the absence of $d=5$ operators
through the implementation of the simplest possible form of the
missing partner
mechanism~\cite{Derendinger:1983aj,Antoniadis:1987dx}. However,
flipped $SU(5)$ gives up one of the most attractive features of
grand unification, namely unification of gauge couplings, because
it is based on the group $SU(5) \times U(1)$. [This is not to say
that the exact unification is impossible within the
four-dimensional flipped $SU(5)$. For the most recent
considerations in this direction see
Refs.~\cite{Ellis:2002vk,Nanopoulos:2002qk}.] Embedding the
flipped $SU(5)$ within an $SO(10)$ gauge group retrieves the gauge
unification but spoils the missing partner mechanism.

The way out, as has been recently shown~\cite{Barr:2002fb}, is to embed the
flipped $SU(5)$ within an $SO(10)$ group in five dimensions using the
extra-dimensional framework {\em \`{a} la\/}
Kawamura~\cite{Kawamura:1999nj,Kawamura:2000ev,Kawamura:2000ir}.
In this way, at the four-dimensional
level, the famous missing partners can still be missing and the doublet-triplet
splitting can be achieved without the dangerous Higgsino-mediated
proton decay. But, one might expect naively that the exact gauge unification is
impossible due to the threshold corrections that originate from the
towers of Kaluza-Klein (KK) modes that are inherent to the
theories with the compactified extra-dimensions.
This naive expectation turns out to be wrong.
The five-dimensional theory, being non-renormalizable, {\em must\/}
have a cutoff ($M_*$). Therefore, the number of KK modes that
contribute is finite. This also makes the threshold corrections finite
and calculable so that the exact unification cannot be excluded {\em a priori}.

This paper is devoted to the issues pertaining to the gauge
coupling unification in the five-dimensional setting. We show,
following the footsteps of Kim and Raby~\cite{Kim:2002im}, that it
is possible to achieve the exact unification using an
$\mathcal{N}=1$ supersymmetric $SO(10)$ model on an $S^1/(Z_2
\times Z_2')$ orbifold. The orbifold has two inequivalent fixed
points, $O$ and $O'$, identified by the action of $(Z_2 \times
Z_2')$ twisting. On the point (brane) $O$ there will be an
$SO(10)$ gauge symmetry while on the point (brane) $O'$ there will
be a flipped $SU(5)$ gauge symmetries. Both symmetries will be the
leftovers of a bigger, $SO(10)$, bulk symmetry. The bulk contains,
besides the vector supermultiplet, a pair of chiral
hypermutiplets: ${\bf 10}_{1H}$ and ${\bf 10}_{2H}$. They give the
Higgs fields of the MSSM: $\overline{{\bf 2}}$ and ${\bf 2}$. The
orbifolding procedure also reduces the amount of the supersymmetry
from $\mathcal{N}=1$ in five dimensions to $\mathcal{N}=1$ in four
dimensions. To obtain the low-energy phenomenology of the Standard
Model (SM) group $\mathcal{H}$ we break flipped $SU(5)$ on the
$O'$ brane by implementing the missing partner mechanism. This
time, in contrast to the model presented in
Ref.~\cite{Barr:2002fb}, we do the breaking with the chiral
superfields that reside on the $O'$ brane.

There are two models in the literature we are going
to compare our results with
that provide the exact gauge coupling unification
in the five-dimensional $S^1/(Z_2 \times Z_2')$ setting.
The common feature for
both models is the placement of the multiplets that contain the Higgs
fields and the gauge sector of the MSSM in the bulk. We briefly review
these models in what follows.
\begin{itemize}
\item The first one is an $SU(5)$ model of Hall and
Nomura~\cite{Hall:2001pg,Hall:2001xb,Hall:2002ci}. In their
model~\cite{Hall:2002ci} the orbifolding yields an $SU(5)$ gauge
symmetry on one brane and the SM gauge symmetry on the other. In
addition, the orbifolding accomplishes the doublet-triplet
splitting by assigning the odd parity to the triplet fields. There
is no need for any extra Higgs breaking except for the usual
electroweak one. For gauge coupling unification not to be spoiled
by arbitrary non-universal contributions coming from the brane
with the SM gauge symmetry Hall and Nomura have to invoke two
requirements: (i) the couplings at the cutoff scale $M_*$ {\em
must\/} enter a strong coupling regime; (ii) the dimension(s) of
the bulk {\em must\/} be large enough (when expressed in terms of
the fundamental scale, i.e.\ cutoff scale, of the theory). We
adopt their requirements in our model, too. \item The second model
is a variant of an $SO(10)$ model of Derm\' \i \v sek and
Mafi~\cite{Dermisek:2001hp}. Here, we just outline the features
that are relevant for comparison with our work. Since the breaking
of $SO(10)$ down to $\mathcal{H}$ demands the reduction of the
group rank \cite{Hebecker:2001jb}, the authors use an extra Higgs
breaking. In the original version of Derm\' \i \v sek and
Mafi~\cite{Dermisek:2001hp} the breaking of $SO(10)$ down to
$SU(5)$ takes place on the $SO(10)$ brane. The low-energy
signature of the SM gauge group is then due to the intersection of
the Pati-Salam and $SU(5)$. The subsequent analysis of the variant
of their model proposed by Kim and Raby~\cite{Kim:2002im}
demonstrated the feasibility of the gauge unification. The
breaking, in Kim and Raby case, takes place on a Pati-Salam brane
affecting only the gauge sector of the theory. [The orbifolding
has already projected out the triplet partners by assigning them
odd parity.] We adopt and extend their method of analysis to
demonstrate the successful unification in our case. The reason
behind the extension is that, in our case, the extra Higgs
breaking affects not only the gauge sector but also the Higgs
sector. Namely, the breaking is what makes the triplets heavy via
missing partner mechanism. This, as it turns out, has significant
consequences on the renormalization group equation (RGE) running
of the gauge couplings as we demonstrate later.
\end{itemize}

In Section~\ref{model} we introduce our model and specify the mass spectrum
of all the fields. We then proceed with
the discussion on the gauge coupling RGE running in five-dimensional orbifold
setting in Section~\ref{KKpart}.
This is where our two main results, the relevant beta coefficients and their RGE
numerical analysis, are presented. Finally, we briefly conclude in
Section~\ref{conclusion}.
\section{An SO(10) model}
\label{model}
We present an $SO(10)$ supersymmetric model in five
dimensions compactified on an $S^1/(Z_2 \times Z'_2)$ orbifold.
The orbifold is created after the fifth dimension, being the circle
$S^1$ of radius $R$, gets compactified through the reflection
$y \rightarrow -y$ under $Z_2$ and $y' \rightarrow -y'$ under
$Z'_2$, where $y'=y+\pi R/2$. There are two fixed points, $O$ and $O'$,
that bound the
physical space $y \in [0,\pi R/2]$ of the bulk. The point $O$ is
referred to as the ``visible brane" while point $O'$ at $y'=0$
is referred to as the ``hidden brane".

We assume that the bulk contains an $\mathcal{N}=1$ vector supermultiplet,
a ${\bf 45}_g$ of $SO(10)$, and two chiral hypermultiplets,
${\bf 10}_{1H}+{\bf 10}_{2H}$. The vector supermultiplet decomposes
into a vector multiplet $V$, which contains the gauge bosons $A_\mu$ and corresponding gauginos, and
a chiral multiplet $\Sigma$ of $\mathcal{N}= 1$ supersymmetry in four dimensions.
Each hypermultiplet splits into two left-handed chiral
multiplets $\Phi$ and $\Phi^c$, having opposite gauge quantum numbers.
To reduce $\mathcal{N}=1$ supersymmetry in
five dimensions to $\mathcal{N}=1$ supersymmetry in four
dimensions we use the parity assignment under $Z_2$. To reduce the gauge
symmetry from $SO(10)$ down to flipped
$SU(5) \otimes U(1)$ on the hidden brane we use the parity assignment
under $Z_2'$. The bulk content of the model is
\begin{subequations}
\label{parity}
\begin{eqnarray}
{\bf 45}_g & = &
V^{++}_{{\bf 24}^0} + V^{++}_{{\bf 1}^0} +
V^{+-}_{{\bf 10}^{-4}} + V^{+-}_{\overline{{\bf 10}}^4} +
\Sigma^{--}_{{\bf 24}^0} + \Sigma^{--}_{{\bf 1}^0} +
\Sigma^{-+}_{{\bf 10}^{-4}} + \Sigma^{-+}_{\overline{{\bf 10}}^4}, \\
{\bf 10}_{1H} & = &
\Phi^{++}_{{\bf 5}_1^{-2}} + \Phi^{+-}_{\overline{{\bf 5}}_1^2} +
\Phi^{c --}_{\overline{{\bf 5}}_1^2} + \Phi^{c-+}_{{\bf 5}_1^{-2}}, \\
{\bf 10}_{2H} & = &
\Phi^{+-}_{{\bf 5}_2^{-2}} + \Phi^{++}_{\overline{{\bf 5}}_2^2} +
\Phi^{c -+}_{\overline{{\bf 5}}_2^2} + \Phi^{c--}_{{\bf 5}_2^{-2}},
\end{eqnarray}
\end{subequations}
where the first (second) superscript denotes the parity assignment under
$Z_2$ ($Z_2'$) transformation.
Only the fields with the $++$ parity contain Kaluza-Klein zero mode fields
($n=0$) that have no effective four-dimensional mass. The masses of all
other modes become quantized
in units of $1/R \equiv M_C$, where $M_C$ is the compactification scale.
For example, all $+-$ and $-+$ parity states are actually the
KK towers of states with
masses $M_C,3M_C,\ldots,(2n+1)M_C,\ldots$, where $n$ is the mode number.

We want to have the low-energy phenomenology that is described by
the SM group $\mathcal{H}$. But, at this point, the brane $O$
feels the $SO(10)$ gauge symmetry while the brane $O'$ feels the
flipped $SU(5)$ gauge symmetry. One could introduce a pair of
Higgses in the bulk, the ${\bf 16}_H$ and the $\overline{\bf
16}_H$, and use the parity assignment to project out all the
states except a pair ${\bf 10}^{1}_H+\overline{{\bf 10}}^{-1}_H$
that is needed for the missing partner mechanism on the visible
brane \cite{Barr:2002fb}. Here, however, we pursue slightly
different direction. Namely, noting that the minimal set of
Higgses that breaks flipped $SU(5)$ down to $\mathcal{H}$ is a
pair of Higgs fields, ${\bf 10}^{1}_H+\overline{{\bf 10}}^{-1}_H$,
we posit their existence on the {\em hidden\/} brane. [Similar
idea on using the minimal Higgs content within an $SU(5)$ model
has been exploited in Ref.~\cite{Hebecker:2001wq}.] With these
fields in place we specify the following brane localized entry of
the superpotential:
\begin{equation}
\label{kappa}
\kappa \Big[\delta \big(y-\frac{\pi R}{2}\big)+\delta \big(y-\frac{3\pi R}{2}\big) \Big]
\Big[\Phi^{++}_{{\bf 5}_1^{-2}}\,{\bf 10}^{1}_H\,{\bf 10}^{1}_H+
\Phi^{++}_{\overline{{\bf 5}}_2^2}\,\overline{{\bf 10}}^{-1}_H\,
\overline{{\bf 10}}^{-1}_H\Big],
\end{equation}
where $\kappa$ represents the Yukawa coupling with the mass
dimension -1/2. Clearly, by giving very large VEVs to the $({\bf
1},{\bf 1},0)$ components of ${\bf 10}^{1}_H$ and $\overline{{\bf
10}}^{-1}_H$, we allow the triplet partners of the doublets in
$\Phi^{++}_{{\bf 5}_1^{-2}}$ and $\Phi^{++}_{\overline{{\bf
5}}_2^2}$ to get large masses through the mating with the triplets
of ${\bf 10}^{1}_H$ and $\overline{{\bf 10}}^{-1}_H$ without
disturbing the lightness of the doublets. This can be
schematically depicted as \cite{Babu:1993we}
\begin{equation}
\label{dts}
\begin{array}{c@{\!\!\!\!\!}c@{\!\!\!\!\!}c|c@{\!\!\!\!\!}c@{\!\!\!\!\!}c}
\left( \begin{array}{c} {\bf 3}\strut\\ \overline{{\bf 2}}\end{array} \right)
& \rule[4.75mm]{16mm}{0.5mm} &
\left( \begin{array}{c} \overline{{\bf 3}}\strut\\ {\rm {\bf other}}\end{array} \right) &
\left( \begin{array}{c} {\bf 3}\strut\\ \overline{{\rm {\bf other}}}\end{array} \right)
& \rule[4.75mm]{16mm}{0.5mm} &
\left( \begin{array}{c} \overline{{\bf 3}}\strut\\ {\bf 2}\end{array}
\right)\strut\\
\parallel & &\parallel & \parallel & &\parallel\strut\\
\Phi^{++}_{{\bf 5}_1^{-2}} & & {\bf 10}^{1}_H & \overline{{\bf
10}}^{-1}_H & & \Phi^{++}_{\overline{{\bf 5}}_2^2}\end{array}
\end{equation}
where, for simplicity, $({\bf 3},{\bf 2}, 1/3) + ({\bf 1},{\bf 1},
0) \equiv {\rm {\bf other}}$, and $\overline{{\bf
3}}=(\overline{{\bf 3}},{\bf 1},2/3)$. Moreover, the symmetry
breaking makes the states $({\bf 1},{\bf 1},0)$, $({\bf 3},{\bf
2},1/3)$, and $(\overline{{\bf 3}},{\bf 2},-1/3)$ from
$V^{++}_{{\bf 24}^0}$ and $V^{++}_{{\bf 1}^0}$ of ${\bf 45}_g$
absorb the corresponding components of the brane Higgses to become
massive, leaving unbroken $\mathcal{H}$ gauge symmetry behind.
[See Table~\ref{t:table2} for the decomposition of $SO(10)$ down
to $\mathcal{H}$ via flipped $SU(5)$.]
\begin{table}
\caption{\label{t:table2} The decomposition of the three lowest lying
representations of $SO(10)$ under the flipped $SU(5)$ group and
the Standard Model gauge group.}
\begin{center}
\begin{tabular}{|c|c|l|}
\hline
$SO(10)$ & $SU(5) \otimes U(1)$ & $SU(3)_c \otimes SU(2)_L \otimes U(1)_Y$\\
\hline
\hline
& ${\bf 24}^0 $ & $({\bf 1},{\bf 1},0) \oplus ({\bf 1},{\bf 3},0) \oplus ({\bf 3},{\bf 2},1/3) \oplus
(\overline{{\bf 3}},\overline{{\bf 2}},-1/3) \oplus ({\bf 8},{\bf 1},0)$ \\
& ${\bf 10}^{-4} $ & $({\bf 1},{\bf 1},-2) \oplus (\overline{{\bf
3}},{\bf 1},-4/3) \oplus
({\bf 3},\overline{{\bf 2}},-5/3)$ \\
\raisebox{2ex}[0pt]{${\bf 45}$}& $\overline{{\bf 10}}^4$ & $({\bf
1},{\bf 1},2) \oplus({\bf 3},{\bf 1},4/3) \oplus
(\overline{{\bf 3}},{\bf 2},5/3)$ \\
& ${\bf 1}^0$ & $({\bf 1},{\bf 1},0)$ \\
\hline
\hline
& ${\bf 1}^{5}$ & $({\bf 1},{\bf 1},2)$ \\
${\bf 16}$ & $\overline{{\bf 5}}^{-3}$ & $({\bf 1},\overline{{\bf 2}},-1) \oplus (\overline{{\bf 3}},{\bf 1},-4/3)$ \\
& ${\bf 10}^1$ & $({\bf 1},{\bf 1},0) \oplus(\overline{{\bf 3}},{\bf 1},2/3) \oplus
({\bf 3},{\bf 2},1/3)$ \\
\hline
\hline
& ${\bf 5}^{-2}$ & $({\bf 1},\overline{{\bf 2}},-1) \oplus ({\bf 3},{\bf 1},-2/3)$ \\
\raisebox{2ex}[0pt]{${\bf 10}$}& $\overline{{\bf 5}}^{2}$ & $({\bf 1},{\bf 2},1) \oplus (\overline{{\bf 3}},{\bf 1},2/3)$ \\
\hline
\end{tabular}
\end{center}
\end{table}

In the discussion from the previous paragraph, we have glossed over a fact
that the bulk fields are KK towers of states. The explicit brane
localized breaking terms will disturb every state of
that tower due to the change of the boundary conditions. Since we want to do
an RGE analysis we need to determine the KK tower position, i.e.\ the mass, of every
state after the disturbance has taken place.
This is what we do next.
\subsection{Mass Spectrum of the Gauge Fields}
\begin{figure}[htb]
\begin{center}
\includegraphics[width=5.5in]{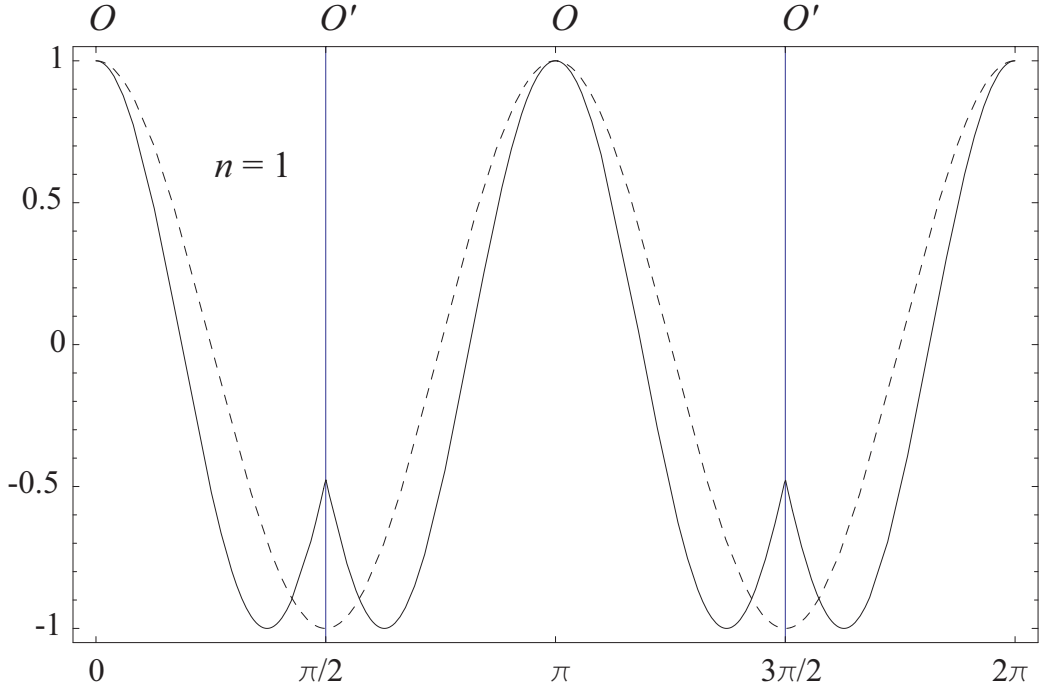}
\end{center}
\caption{\label{deltaKK} A plot of an $n=1$ mode of the bulk field
wave function profile in
the fifth dimension. The dashed line represents undisturbed profile
given by $\cos 2ny$. The solid line
represents the profile after the perturbation due to the
boundary condition is accounted for. The radius $R$ is taken to be 1.}
\end{figure}

The five-dimensional theory is non-renormalizable. Therefore, we expect the
theory to have a cutoff scale $M_*$ where some new physics comes into
play (e.g.\ other dimensions beyond five, strings). We take the VEVs of
the symmetry breaking Higgs fields to be of the order of this cutoff:
$\langle ({\bf 1},{\bf 1},0) \rangle \equiv M \sim M_*$. Then the
Lagrangian involving the gauge fields gets additional contribution
\cite{Nomura:2001mf,Kim:2002im}
\begin{equation}
\mathcal{L} \subset \frac{1}{2}
\Big[\delta \big(y-\frac{\pi R}{2}\big)+\delta \big(y-\frac{3\pi R}{2}\big) \Big]
g_5^2 M^2 A^{\hat{a}}_{\mu} A^{\hat{a} \mu},
\end{equation}
where $g_5^2$ represents the gauge coupling of the
five-dimensional theory and $\hat{a}$ is an $SO(10)$ group index
that goes through all the gauge fields associated with the broken
$++$ parity generators we mentioned at the end of
Section~\ref{model}. [The five-dimensional gauge coupling $g_5^2$
has mass dimension $-1$.] The equations of motion for the
``broken" gauge bosons are \cite{Nomura:2001mf,Kim:2002im}
\begin{equation}
\label{deltapotential}
-\partial_y^2 A^{\hat{a}}_{\mu}(x,y) +
\Big[\delta \big(y-\frac{\pi R}{2}\big)+\delta \big(y-\frac{3\pi R}{2}\big) \Big]
g_5^2 M^2 A^{\hat{a}}_{\mu} (x,y)= (M^A_n)^2 A^{\hat{a}}_{\mu}(x,y),
\end{equation}
where $M^A_n$ represents the effective Kaluza-Klein mass in four
dimensions of the $n$th mode. It is defined via Klein-Gordon
equation $\big[\partial_\nu
\partial^\nu+(M^A_n)^2\big]A^{\hat{a}}_{\mu}(x,y)=0$. The second
term on the left-hand side of Eq.~\eqref{deltapotential} is
responsible for the deviation from the usual mass spectrum of the
$++$ parity fields ($M^A_n=0, 2 M_C,\ldots,2 n M_C,\ldots$). It
reminds us of the delta function-type potential in ordinary
Schr\"{o}dinger's equation. The role of this term is thus to repel
the bulk field wave function away from the brane. In the language
of the effective four-dimensional theory this means that even the
zero mode ($n=0$) of the gauge bosons becomes massive. Taking the
following ansatz for the five-dimensional gauge field on the
segment $y \in [0,\pi R/2]$:
\begin{equation}
A^{\hat{a}}_{\mu} (x,y)=\frac{1}{\sqrt{\pi R}}
\sum^{\infty}_{n=0} N_n A^{\hat{a}(n)}_\mu (x) \cos M_n^Ay,
\end{equation}
the eigenvalue equation for the effective mass, due to the nontrivial
boundary condition at the hidden brane, takes the form \cite{Nomura:2001mf}
\begin{equation}
\label{conditionG}
\tan \frac{M_n^A \pi R}{2} =\frac{g_5^2 M^2}{2 M_n^A}.
\end{equation}
The normalization constant for the $++$ parity bulk fields also changes from
$1/\sqrt{2^{\delta_{n0}}}$ to $N_n=\big[1+M_C g_5^2 M^2
\cos^2 \frac{M_n^A \pi R}{2}/(\pi (M_n^A)^2)\big]^{-1/2}$ \cite{Choi:2003bh}.
The plot of the modified wave function profile for $n=1$ is given in Fig.~\ref{deltaKK}.
[We excluded the normalization constants for simplicity.]

There are two interesting approximations that we can consider:
$g_5^2 M^2 \gg M_n^A$ and $g_5^2 M^2 \ll M_n^A$. The former one
generates the following approximate solution of the eigenvalue
equation for the mass spectrum
\begin{equation}
M_n^A \simeq (2n+1) M_C \big[1-\varepsilon+\varepsilon^2\big],
\end{equation}
while the latter one yields
\begin{equation}
M_0^A \simeq 2 M_C \sqrt{\frac{1}{\pi^2 \varepsilon}},
\;\;\mbox{and}\;\;
M_{n \neq 0}^A \simeq 2n M_C \big[1+\frac{1}{\pi^2 \varepsilon n^2}\big],
\end{equation}
where we define $\varepsilon \equiv (4 M_C)/(\pi g_5^2 M^2)$.
The two approximations generate qualitatively different mass spectra. Therefore,
it is very important to determine which one is applicable to our scenario.
Assuming that all the couplings of the theory enter the strong regime
at the cutoff $M_*$ we can use the result of the naive dimensional
analysis~\cite{Chacko:1999hg} in higher dimensional theories that suggests
$g_5^2 \simeq 24 \pi^3/M_*$ and $M \simeq M_*/(4 \pi)$, which gives
$g_5^2 M^2 \simeq 3/2 \pi M_* > M_* \gg M_n^A$.
We thus choose the former approximation. Following the work of Kim and
Raby~\cite{Kim:2002im}, we introduce the parameter $\zeta = 2 N \varepsilon$,
where $2N=M_*/M_C$ and $\zeta \simeq 8/(3 \pi^2) \simeq 0.27$, to rewrite
the approximate mass spectrum of the broken gauge bosons as
\begin{equation}
M_n^A \simeq M_C \big(2n+1-\frac{n}{N} \zeta \big).
\end{equation}

One interesting feature to note is that the boundary condition in
Eq.~\eqref{conditionG} is not absolute \cite{Nomura:2001mf}. In
our case, the broken $++$ parity field modes start off with the
mass spectrum that mimics the spectrum of the $+-$ and $-+$ parity
field modes but then gradually merges with the spectrum of
undisturbed $++$ and $--$ parity bulk fields as one moves up the
Kaluza-Klein tower of states. One should also keep in mind that
the supersymmetry ensures the same fate for the chiral partners
$\Sigma$ of the vector fields $V$. Namely, the mass spectrum of
the fields in $\Sigma^{--}$ are shifted in the same manner as the
states in the $V^{++}$ that are made massive through the brane
gauge breaking. With that said, we turn to the consideration of
the Higgs field mass spectrum.
\subsection{Mass Spectrum of the Higgs Fields}
The missing partner mechanism affects {\em only\/} the color
triplets of the bulk states with $++$ and $--$ parities. To
determine their effective mass spectrum we concentrate on the
masses of the color Higgsinos. Supersymmetry then ensures the same
mass spectrum for their bosonic partners. Moreover, since there
are two separate color triplet sectors, as indicated by the
vertical line in Eq.~\eqref{dts}, we treat only one of them. The
other sector will have the same mass spectrum as long as both
sectors share the {\em same\/} dimensionful coupling $\kappa$. We
assume this to be the case. Note that the bulk states with the
$+-$ and $-+$ parities, i.e.\ the odd states, do not get affected
by the brane breaking.

To make the discussion as transparent as possible we adopt the
following notation for the triplet Higgsinos: $H_C \in
\Phi^{++}_{{\bf 5}_1^{-2}}$, $H_{C}^c \in
\Phi^{c--}_{\overline{{\bf 5}}_1^{2}}$, and $H_{\overline{C}_H}
\in {\bf 10}^{1}_H$. Their equations of motion, derived from the
brane coupling term in Eq.~\eqref{kappa} and the bulk action (see
\cite{Arkani-Hamed:2001tb}), read \cite{Choi:2003bh}
\begin{align}
&{\rm i}{\bar\sigma}^\mu\partial_\mu H_{\overline{C}_H}-\kappa M \overline{H}_{C}
|_{y=(\pi R/2,\,3 \pi R/2)}=0, \\
&{\rm i}{\bar\sigma}^\mu\partial_\mu H_{C}-\partial_y \overline{H}^c_{C}
-\kappa M \overline{H}_{\overline{C}_H} \big(\delta(y-\pi R/2)+\delta(y-3\pi R/2)\big)=0,\\
&{\rm i}{\bar\sigma}^\mu\partial_\mu H^c_{C}
+\partial_y \overline{H}_{C}=0.
\end{align}
These equations are satisfied by the following ansatz for the five-dimensional
Higgsino fields on the segment $y \in [0,\pi R/2]$
\begin{align}
H_{C}(x,y)  &=\frac{1}{\sqrt{\pi R}}\sum_n N^{H_C}_n h^{(n)}_1(x) \cos M^{H_C}_n y,\\
H^c_{C}(x,y)&=\frac{1}{\sqrt{\pi R}}\sum_n N^{H_C}_n h^{(n)}_2(x) \sin M^{H_C}_n y,\\
\intertext{and the Higgsino field localized on the hidden brane}
H_{\overline{C}_H}(x)&=\frac{1}{\sqrt{\pi R}}\sum_n N^{H_C}_n
\frac{\kappa M}{M^{H_C}_n} h^{(n)}_2(x) \cos \frac{M^{H_C}_n \pi
R}{2}.
\end{align}
Here, the eigenvalue equation for the effective mass, due to the nontrivial
boundary condition at the hidden brane, takes the form \cite{Choi:2003bh}
\begin{equation}
\label{conditionH}
\tan \frac{M_n^{H_C} \pi R}{2} =\frac{\kappa^2 M^2}{2 M_n^{H_C}},
\end{equation}
where we define the effective KK mass via a pair of Weyl equations:
${\rm i}{\bar\sigma}^\mu\partial_\mu h^{(n)}_1=M^{H_C}_n
{\overline h}^{(n)}_2$ and ${\rm i}{\bar\sigma}^\mu\partial_\mu h^{(n)}_2=M^{H_C}_n
{\overline h}^{(n)}_1$.

The naive dimensional analysis~\cite{Chacko:1999hg} in the strong
coupling regime yields $\kappa \simeq (24 \pi^3/M_*)^{1/2}$, which implies
that $\kappa^2 M^2 (\simeq g_5^2 M^2) \gg M^{H_C}_n$. In this limit, the mass
spectrum of the Higgsino triplets looks, in form, exactly the same as the mass
spectrum of the broken gauge fields. Namely, the mass eigenvalues of
Eq.~\eqref{conditionH} are
\begin{equation}
M_n^{H_C} \simeq M_C \big(2n+1-\frac{n}{N} \zeta \big),
\end{equation}
where we assume that $\kappa^2 M^2 = g_5^2 M^2$ for simplicity.
For completeness, the normalization constant $N_n^{H_C}$ is \cite{Choi:2003bh}
\begin{equation}
N^{H_C}_n=\bigg(1+\frac{M_C \kappa^2 M^2}{\pi(M^{H_C}_n)^2}
\cos^2\frac{M^{H_C}_n\pi R}{2}\bigg)^{-1/2}.
\end{equation}

\begin{figure}[htb]
\begin{center}
\includegraphics[width=6in]{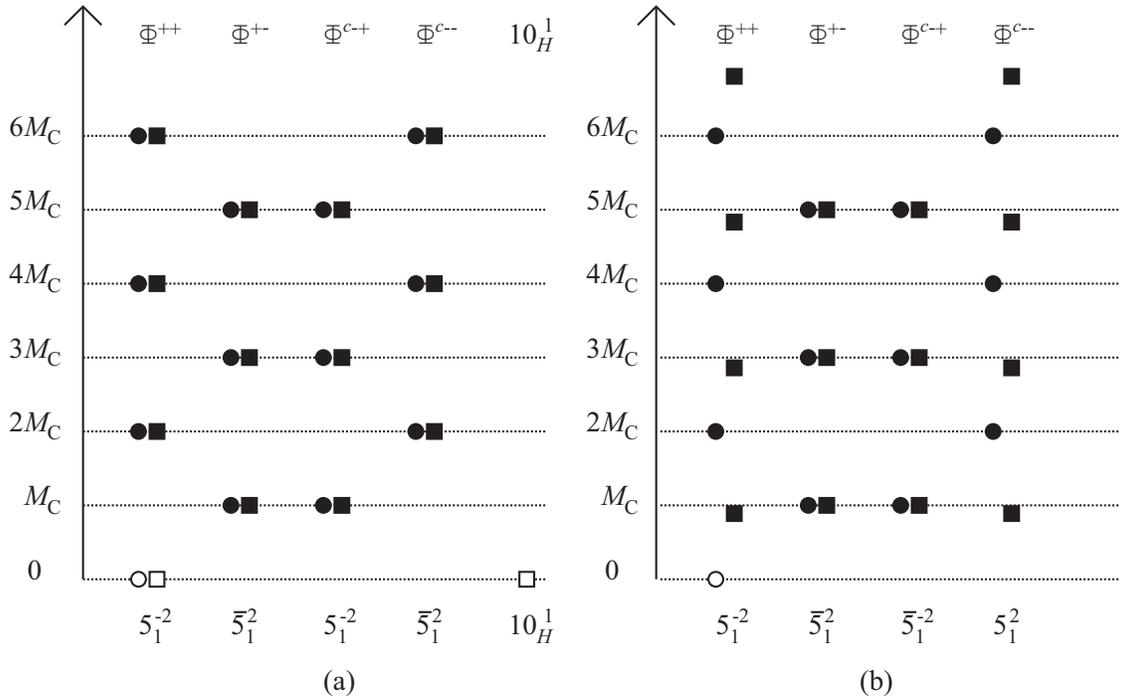}
\end{center}
\caption{\label{higgssector} (a) A mass spectrum of the Kaluza-Klein towers of
the Higgs sector after the compactification, but before the brane localized breaking. (b)
The mass spectrum after the brane localized breaking. The circles represent the doublets
and the squares represent the triplets.}
\end{figure}

In the case of the color Higgsinos there is a mixing between the bulk and the
brane fields. It is the role of the brane field $H_{\overline{C}_H}$ to give the
mass to the zero mode component of $H_C$. As described in Ref.~\cite{Choi:2003bh},
the Weyl spinors, $h_1^{(n)}$ and $h_2^{(n)}$, pair up at every Kaluza-Klein level
to obtain the Dirac mass. The remaining states in the ${\bf 10}^{1}_H$ of Higgs get
absorbed by the broken gauge bosons and completely disappear as far as the
running is concerned.
We show the mass spectrum of one part of the Higgs sector in Fig.~\ref{higgssector}.
The other part looks exactly the same.
Since this concludes the discussion on the mass spectrum of
both the gauge and the Higgs fields we turn our attention towards the RGE
analysis.
\section{Kaluza-Klein unification}
\label{KKpart}
The running of the gauge couplings in our model is the same as the running
in the usual four-dimensional theory as long as we stay below the compactification
scale $M_C$. But, once we venture over $M_C$, the running is affected by the
towers of Kaluza-Klein states until we reach the cutoff scale $M_*$, which we
define as the scale where effective gauge couplings merge. Since there are
numerous states in the KK towers one might expect that the analysis of the
threshold effects on the gauge coupling running from $M_C$ to $M_*$ is very
difficult even at a one-loop level. This, however, is not the case as we show next.

Let us, for concreteness, limit our discussion to the
five-dimensional theory that is based on the simple gauge group
$\mathcal{F}$. The main simplification originates from the
observation that the compactification procedure forces all the
states that make up a single representation of $\mathcal{F}$ to
appear within the interval $[2nM_C,2(n+1)M_C]$ for every $n \neq
0$. [This statement is true regardless of the type of the
additional brane boundary conditions we discussed in the previous
two sections.] These states obviously contribute in an
$\mathcal{F}$ invariant way to the running of all the gauge
coupling constants after we go over $2(n+1)M_C$. Thus, the
contribution of the $n$th Kaluza-Klein level that starts to appear
at $2nM_C$ drops out of the running of the difference of the gauge
couplings after we reach $2(n+1)M_C$. In view of this fact we are
motivated to pursue the differential running, i.e.\ the running of
the difference of the gauge couplings. The previous observation
also implies that the beta coefficients reset themselves to the
values of the familiar coefficients of the Standard Model group
$\mathcal{H}$ every time we go over another $2(n+1)M_C$ scale.

Nontrivial boundary conditions distort the spectrum of Kaluza-Klein masses. In
our case, the members of the $n$th mode emerge at
$2nM_C$, $(2n+1-\frac{n}{N} \zeta)M_C$, $(2n+1)M_C$, and $(2n+2)M_C$ energy
levels.
We have already concluded that from $2nM_C$ to $(2n+1-\frac{n}{N} \zeta)M_C$
the beta coefficients {\em must\/} be the coefficients of the SM group $\mathcal{H}$.
We call this region I.
Region II is the region from $(2n+1-\frac{n}{N} \zeta)M_C$ to $(2n+1)M_C$, while
region III stretches from $(2n+1)M_C$ to $(2n+2)M_C$ for $n \neq 0$. The notation
here and in what follows is exactly the same as the notation of Kim and
Raby~\cite{Kim:2002im}. Note that we do not mention the matter fields at any
point. The reason is that the matter fields of one family contribute equally to
the running of the gauge couplings regardless of their origin, i.e.\ whether they
are located in the bulk or on the brane.

As shown by Kim and Raby~\cite{Kim:2002im}, if the
compactification breaks $\mathcal{F}$ to $\mathcal{G}$ and, then,
the brane breaking reduces $\mathcal{G}$ to the SM group
$\mathcal{H}$, the beta coefficients of the gauge sector are:
\begin{equation}
\label{b:gauge}
\begin{split}
b^{\rm I}_{\rm gauge} & = b^\mathcal{H}(V); \\
b^{\rm II}_{\rm gauge} & = b^\mathcal{H}(V) + b^\mathcal{G/H}(V) + b^\mathcal{G/H}(\Sigma)
= b^\mathcal{G}(V) + b^\mathcal{G}(\Sigma) - b^\mathcal{H}(\Sigma);\\
b^{\rm III}_{\rm gauge} & = b^\mathcal{H}(V) + b^\mathcal{G/H}(V) + b^\mathcal{G/H}(\Sigma) +
b^\mathcal{F/G}(V)+ b^\mathcal{F/G}(\Sigma)
= -b^\mathcal{H}(\Sigma).
\end{split}
\end{equation}
[The notation is that $b \equiv(b_1,b_2,b_3)$, where $b_1$, $b_2$,
and $b_3$ are the coefficients associated with the gauge couplings
of $U(1)_Y$, $SU(2)_L$, and $SU(3)_c$ respectively.] Here, we use
the fact that $b^\mathcal{F}(V)\equiv b^\mathcal{H}(V) +
b^\mathcal{G/H}(V) + b^\mathcal{F/G}(V)$ is an $\mathcal{F}$
invariant coefficient that drops out from the running of the
differences of the gauge couplings. The same statement holds for
$b^\mathcal{F}(\Sigma)\equiv b^\mathcal{H}(\Sigma) +
b^\mathcal{G/H}(\Sigma) + b^\mathcal{F/G}(\Sigma)$ coefficient.
$\mathcal{G/H}$ and $\mathcal{F/G}$ represent the appropriate
coset-spaces (e.q.\ states that are in $\mathcal{G} \supset
\mathcal{H}$ but {\em not\/} in $\mathcal{H}$ belong to
$\mathcal{G/H}$). Note that we always have $b(\Sigma)=-b(V)/3$
since $\Sigma$ is the chiral superfield and $V$ is the vector
superfield. In our case $\mathcal{F}$ corresponds to $SO(10)$ and
$\mathcal{G}$ corresponds to the flipped $SU(5)$ group.

Before we consider the beta coefficient of the Higgs sector we
note the following: the beta coefficients of the {\em two\/}
supersymmetric Higgs doublets (triplets) are $b({\bf 2}) \equiv
(3/5,1,0)$ ($b({\bf 3}) \equiv (2/5,0,1)$). Therefore, the sum of
the contributions of the pair of doublets and the pair of triplets
does not affect the differential running and can be freely
discarded. Moreover, as far as the differential running is
concerned, we can write $b({\bf 2})=-b({\bf 3})=(0,2/5,-3/5)$,
where we subtract the overall constant to make $b_1=0$. This we do
with all the other beta coefficients in what follows. Recalling
that there are two Higgs sectors we can write:
\begin{equation}
\label{b:Higgs}
\begin{split}
b^{\rm I}_{\rm Higgs} & = b({\bf 2}); \\
b^{\rm II}_{\rm Higgs} & = b({\bf 2})+2 b({\bf 3}) = b({\bf 3});\\
b^{\rm III}_{\rm Higgs} & = b({\bf 3})+2 b({\bf 2})+2 b({\bf 3})= b({\bf 3}).
\end{split}
\end{equation}

Finally, we are ready to analyze the running at one-loop level. The relevant
RGEs and all the definitions are taken from Kim and Raby~\cite{Kim:2002im}. We
present them here for completeness of this work. The one-loop RGEs for the gauge
couplings in the effective four-dimensional theory are
\begin{equation}
\label{RGE:5D}
\frac{2 \pi}{\alpha_i(\mu)}=\frac{2 \pi}{\alpha(M_*)}+
\big[b^{\mathcal{H}}_i(V)+
b^{\mathcal{H}}_i({\bf 2})+b^{\mathcal{H}}_{\rm matter}\big] \ln \frac{M_C}{\mu}+
\Delta^{\rm Higgs}_i+\Delta^{\rm gauge}_i,
\end{equation}
where $\Delta$'s describe the appropriate threshold corrections of the
Kaluza-Klein modes from $M_C$ to $M_*$. They are given by
\begin{equation}
\Delta \equiv \ b^{\rm eff} \ln \frac{M_*}{M_C} = b^{\rm I} A_{\rm I}
+ b^{\rm II} A_{\rm II} + b^{\rm III} A_{\rm III},
\end{equation}
with
\begin{subequations}
\begin{eqnarray}
A_{\rm I} & = & \sum_{n=1}^{N-1} \ln \frac{2n+1 -\frac{n}{N} \zeta}{2n}, \\
A_{\rm II} & = & \sum_{n=1}^{N-1} \ln \frac{2n+1}{2n+1- \frac{n}{N} \zeta }, \\
A_{\rm III} & = & \sum_{n=1}^N \ln \frac{2n}{2n-1}.
\end{eqnarray}
\end{subequations}
Obviously, $A_{\rm I}$, $A_{\rm II}$ and $A_{\rm III}$ allow us to sum over
the threshold corrections from the corresponding regions.

Taking the large $N$ limit, where $2N = M_*/M_C$, and using the
approximation $\ln (1+x) = x + \cdots$, Kim and
Raby~\cite{Kim:2002im} give the following expression for the
threshold corrections of the gauge and the Higgs sector:
\begin{equation}
\Delta = \frac{1}{2}(b^{\rm III}+b^I) \ln \frac{M_*}{M_C} + \frac{1}{2}
(b^{\rm III} - b^{\rm I}) \ln \frac{\pi}{2} + \frac{1}{2} (b^{\rm II}-b^{\rm I}) \zeta.
\end{equation}
Looking back at Eqs.~\eqref{b:gauge} and \eqref{b:Higgs} we have
for our model
\begin{subequations}
\begin{eqnarray}
\Delta^{\rm gauge} & = & \frac{2}{3} b^{\mathcal{H}}(V)\ln\frac{M_*}{M_C}
-\frac{1}{3}b^{\mathcal{H}}(V)\ln\frac{\pi}{2}+
\frac{1}{3}\big[b^{\mathcal{G}}(V)-b^{\mathcal{H}}(V)\big] \zeta,\label{gaugecorrection}\\
\Delta^{\rm Higgs} & = & -b({\bf 2}) \ln\frac{\pi}{2} - b({\bf 2}) \zeta.\label{Higgscorrection}
\end{eqnarray}
\end{subequations}
Moreover, since $b^{\mathcal{H}}(V)$ represents the beta
coefficients of the gauge sector of the MSSM we have
$b^{\mathcal{H}}(V)=(0,-6,-9)$. On the other hand,
$b^{\mathcal{G}}(V)$ represents the beta coefficients of the gauge
sector of the supersymmetric flipped $SU(5)$: ${\bf 24}^0+{\bf
1}^0$. Therefore, $b^{\mathcal{G}}(V)=(-3/5,-15,-15) \equiv
(0,-72/5,-72/5)$, where we again subtract the overall constant
contribution to make $b_1$ coefficient equal to zero. Using these
results we find:
\begin{subequations}
\begin{eqnarray}
\Delta^{\rm gauge} & = & \big(0,-4 \ln\frac{M_*}{M_C}
+2 \ln\frac{\pi}{2}-\frac{14}{5} \zeta,
-6 \ln\frac{M_*}{M_C}
+3 \ln\frac{\pi}{2}-\frac{9}{5} \zeta\big),\\
\Delta^{\rm Higgs} & = & \big(0,-\frac{2}{5} \ln \frac{\pi}{2}-\frac{2}{5} \zeta
,\frac{3}{5} \ln \frac{\pi}{2}+\frac{3}{5} \zeta \big).
\end{eqnarray}
\end{subequations}

\begin{figure}[htb]
\begin{center}
\includegraphics[width=4.5in]{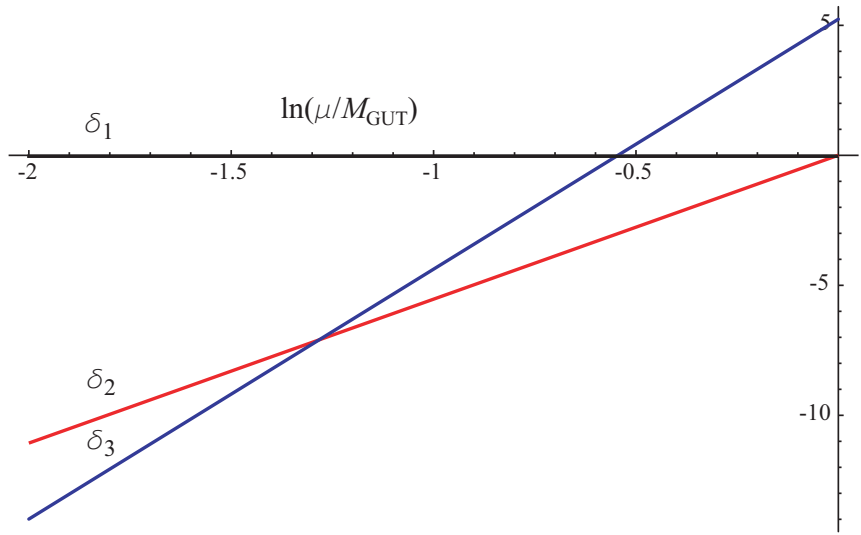}
\end{center}
\caption{\label{alphainvdiffplot} A plot of the differential running
$\delta_i(\mu)=2 \pi (1/\alpha_i(\mu)-1/\alpha_1(\mu))$ versus
$\ln (\mu/M_{\rm GUT})$, where $M_{\rm GUT} = 2.37 \times 10^{16}$\,GeV.}
\end{figure}

Our goal is to find the values of $M_C$ and $M_*$ that allow the
exact unification, at least at one-loop level, of the gauge
coupling constants at the scale $M_*$. To be able to do that we
first recall the situation we have in the usual four-dimensional
SUSY GUT. There we define $M_{\rm GUT}$ to be the scale where
$\alpha_1(M_{\rm GUT})=\alpha_2(M_{\rm GUT}) \equiv
\tilde{\alpha}_{\rm GUT}$ \cite{Kim:2002im} with the running given
by
\begin{equation}
\label{RGE:4D}
\frac{2 \pi}{\alpha_i(\mu)}=\frac{2 \pi}{\alpha_i(M_{\rm GUT})}+\big[b^{\mathcal{H}}_i(V)+
b^{\mathcal{H}}_i({\bf 2})+b^{\mathcal{H}}_{\rm matter}\big]
\ln \frac{M_{\rm GUT}}{\mu}.
\end{equation}
If we ask how far off from $\tilde{\alpha}_{\rm GUT}$ the coupling
$\alpha_3(M_{\rm GUT})$ is, and parameterize the degree of
nonunification via $\delta_3=\big(2 \pi/\alpha_3(M_{\rm GUT})-2
\pi/\tilde{\alpha}_{\rm GUT}\big)$, we obtain $5 \lesssim \delta_3
\lesssim 6$ depending on the exact spectrum of SUSY particles. We
show one example of differential running in
Fig.~\ref{alphainvdiffplot}. This example takes into the account
not only the one-loop but the two-loop effects on the running of
the gauge couplings. We also assume that the superpartners have
masses of the order of $m_t$, and take the lower experimental
limit $\tan \beta=3$ \cite{:2001xx}.

In the five-dimensional setting the deviation from the usual
running starts at $M_C$ scale. Therefore, at $M_C$, the left-hand
sides of Eqs.~\eqref{RGE:5D} and \eqref{RGE:4D} {\em must\/} be
the same. Thus, we have that
\begin{equation}
\begin{split}
\label{xxxx}
\delta_2(M_C)&=\big[b^{\mathcal{H}}_2(V)+
b^{\mathcal{H}}_2({\bf 2})]\ln \frac{M_{\rm GUT}}{M_C}=\Delta^{\rm gauge}_2+\Delta^{\rm Higgs}_2,\\
\delta_3(M_C)&= \big(2 \pi/\alpha_3(M_{\rm GUT})-2
\pi/\tilde{\alpha}_{\rm GUT}\big)\\
&\quad +\big[b^{\mathcal{H}}_3(V)+ b^{\mathcal{H}}_3({\bf 2})]\ln
\frac{M_{\rm GUT}}{M_C}=\Delta^{\rm gauge}_3+\Delta^{\rm Higgs}_3.
\end{split}
\end{equation}
Solving these equations yields
\begin{equation}
M_C \approx 5.5 \times 10^{14}\,{\rm GeV},\,\,\mbox{and}\,\,
M_* \approx 1.0 \times 10^{17}\,{\rm GeV},
\end{equation}
where we use the same value of $\delta_3$ as is used by Kim and
Raby~\cite{Kim:2002im} ($\delta_3 \simeq 6$) and we take the
corresponding value of $M_{\rm GUT}$ ($M_{\rm GUT}=3 \times
10^{16}$\,GeV). These values imply that $N=90$, justifying the
large $N$ approximations. This also ensures that the effect of the
non-universal brane kinetic terms, present on the $O'$ brane, on
the gauge coupling unification is sufficiently small to be
neglected~\cite{Hall:2001pg}.

In view of our results
the following picture emerges. The effective theory below the compactification scale
looks exactly the same as the usual MSSM theory. Then, once we go above $M_C$, there
emerge the towers of the Kaluza-Klein states that change the behavior of the gauge
running through the set of small but numerous threshold corrections. The theory finally
yields the gauge unification at $M_* > M_{\rm GUT}$ where all the couplings of the
theory enter the strong regime. At that point the five-dimensional theory must be
embedded into more fundamental physical picture.

We should note that our result is not very sensitive to the exact
value of the small parameter $\zeta$. On the other hand, the
values of $M_C$ and $M_*$ depend very strongly on the value of
$\delta_3$. We have taken $\delta_3 \simeq 6$ to be able to
compare our results with the analysis of Kim and
Raby~\cite{Kim:2002im}. This value, coming from the RGE
propagation of the experimental value of $\alpha_3(m_Z)=0.118 \pm
0.003$ \cite{Hagiwara:fs} from the $m_Z$ scale to the GUT scale,
could be reduced in near future. Namely, the new estimate of
$\alpha_3$ from $\tau$ lifetime suggests
$\alpha_3(m_Z)=0.1221^{+0.0026}_{-0.0023}$
\cite{Erler:2002bu,Langacker:2003tv}. This would have a large
impact on our result since the corresponding value of $\delta_3$
($\delta_3 \simeq 3$) would imply $N=2$, making the whole KK
unification picture questionable. The model of Kim and
Raby~\cite{Kim:2002im} for the case of $\delta_3 \simeq 3$ yields
$N=27$.

This paper is devoted solely to the analysis of the gauge coupling
unification. This means that there are many questions left
unanswered. For example, one might ask what mechanism breaks
four-dimensional $\mathcal{N}=1$ supersymmetry. Or, how the Higgs
fields responsible for the missing partner mechanism get their
VEVs. Our intention was not to answer the questions like these but
to demonstrate the possibility of the five-dimensional
Kaluza-Klein unification and this we did. But, some of these
questions, including the possibility of having a model with the
realistic mass patterns, have already been tackled in
Ref.~\cite{Barr:2002fb}. [There are, of course, different
directions one might take. Namely, a number of five-dimensional
$SO(10)$ models with the Pati-Salam signature on one brane and
$SO(10)$ signature on the other brane has been studied in the
literature
\cite{Dermisek:2001hp,Albright:2002pt,Kyae:2002ss,Kim:2002im,Kim:2003vr}.
Even the most general scenario of having a model with the
five-dimensional $SO(10)$ gauge symmetry that is broken by
compactification on both branes has also been investigated
recently \cite{Kyae:2003ek}.]

Our result for $M_C$ and $M_*$ is very similar to the result
obtained by Kim and Raby~\cite{Kim:2002im}. This is due to the
fact that the biggest correction to the standard four-dimensional
running in both cases comes from the first term in
Eq.~\eqref{gaugecorrection}. Since this term involves the beta
coefficients of the SM gauge group only, the leading corrections
must be the same for all the schemes with the realistic low-energy
signature. The main difference between the two models in the gauge
sector is generated by the beta coefficients $b^{\mathcal{G}}(V)$
of the gauge group on the hidden brane. In our case the hidden
brane has the flipped $SU(5)$ group with
$b^{\mathcal{G}}(V)=(0,-72/5,-72/5)$, while in the case of Kim and
Raby the hidden brane harbors PS gauge group with
$b^{\mathcal{G}}(V)=(0,12/5,-18/5)$. The main difference in the
Higgs sector stems from the fact that there is no distinction
between the region I and region II in Kim and Raby case since the
additional boundary conditions do not affect the Higgs sector at
all. Therefore, the second term in Eq.~\eqref{Higgscorrection} is
absent in their case. It is interesting to note that the
difference between the two models is in the terms that are
proportional to the small parameter $\zeta$. Therefore, the limit
$\zeta \rightarrow 0$ gives the same result in both cases. In that
limit we obtain $M_C \approx 3.2 \times 10^{14}\,{\rm GeV}$, and
$M_* \approx 2.2 \times 10^{17}\,{\rm GeV}$. Interestingly enough,
the same limit reproduces the results of the analysis on the gauge
coupling unification of the five-dimensional $SU(5)$ model
\cite{Hall:2002ci}. One can even make a more general
statement\footnote{We thank Hyung Do Kim for pointing this out to
us.} about various models yielding the same result in the limit
when the brane breaking is large enough ($\zeta \rightarrow 0$).
Namely, one expects the same corrections to the usual
four-dimensional running in all models that fulfill the following
conditions: i) $\mathcal{F}$ is a unified group; ii) $\mathcal{H}$
corresponds to the SM group; iii) Symmetry breaking $\mathcal{G}
\rightarrow \mathcal{H}$ is localized at the $\mathcal{G}$ brane;
iv) the MSSM Higgses originate from the bulk. Clearly, all of the
above conditions are satisfied by the models we consider.

Even though the exact unification of the gauge couplings in the
four-di\-men\-sional flipped $SU(5)$ cannot be excluded
\cite{Ellis:2002vk,Nanopoulos:2002qk}, one can never justify the
charge quantization and the hypercharge assignment without
embedding it into $SO(10)$. In our case this is not an issue. As
long as the matter fields are placed in the bulk or on the visible
brane we guarantee the charge quantization. [Of course, if the
matter comes from the bulk multiplets we might lose the
unification of quarks and leptons of one family.] The exact
location of the matter fields is conditioned by the presence of
$d=6$ proton-decay operators induced by the exchange of the $X$
gauge bosons. Namely, the experimental limit on proton lifetime
yields the limit of $M > 2.8 \times 10^{15}\,{\rm GeV}$ on the
mass $M$ of the $X$ gauge bosons within the four-dimensional
flipped $SU(5)$ \cite{Murayama:2001ur}. Since the mass spectrum of
$X$ bosons in our model starts from the compactification scale
($M_C \approx 6 \times 10^{14}\,{\rm GeV}$) it is clear that not
all the families of matter fields can be placed on the visible
brane. It is necessary for, at least, the first and the second
family to come from the bulk multiplets. The idea of localizing
the matter fields on the flipped $SU(5)$ brane does not appeal to
us on the grounds of charge quantization. But, in that case, the
suppression of the gauge field wave function on the flipped
$SU(5)$ brane that is visible in Fig.~\ref{deltaKK} is sufficient
to make the prediction for the proton decay via $p \rightarrow
\pi^0 e^+$ channel very close to the present experimental bound
(see for example \cite{Nomura:2001mf,Hebecker:2002rc}). In this
aspect the model of Kim and Raby~\cite{Kim:2002im} does better job
since the localization of the matter fields on the PS brane, in
their case, still justifies the charge quantization. The only {\em
ad hoc\/} feature of our model is the existence of the Higgses on
the hidden brane. It is difficult to justify their $U(1)$ charges
unless they originate from the ${\bf 16}$ and the $\overline{{\bf
16}}$ bulk fields. [This remains an open possibility.] We argue
that their $U(1)$ charges are what one expects from the fields of
flipped $SU(5)$ and that they provide the anomaly cancellation on
the hidden brane. The model can still produce interesting mass
matrix patterns $L = D$ and $N = U$ that were discussed in
Ref.~\cite{Barr:2002fb} where, in our case, the relation $L = D$
holds only for the third family. In addition, it has been shown
that this class of models allows for gaugino meditated
supersymmetry breaking with the non-universal gaugino
masses~\cite{Barr:2002fb} which leads to the realistic
supersymmetry mass spectra~\cite{Baer:2002by,Balazs:2003mm}.
\section{Conclusion}
\label{conclusion}
We have presented an $SO(10)$ model in five dimensions. The model
has served to demonstrate that the exact unification of the gauge couplings
is possible even in the higher dimensional setting. The corrections to the
usual four-dimensional running have been due to the Kaluza-Klein towers
of states. We have shown that despite the large amount of these states
the corrections for the MSSM running can be unambiguously and systematically
evaluated. Demanding the exact unification, the compactification scale is deduced
to be $M_C \approx 5.5 \times 10^{14}\,{\rm GeV}$ with the cutoff of the theory at
$M_* \approx 1.0 \times 10^{17}\,{\rm GeV}$. Therefore, the five-dimensional theory
exists in a rather large energy region before one needs to replace it with the more
fundamental one.

The usual problems of SUSY GUTs, such as the doublet-triplet
splitting problem, have been solved in a natural way. For example,
the presence of the flipped $SU(5)$ symmetry on the hidden brane
has allowed us to implement the missing partner mechanism. At the
same time the presence of the $SO(10)$ symmetry on the visible
brane still allows one to obtain desirable predictions for the
quark and lepton masses such as $m_b = m_{\tau}$. The model yields
the low-energy signature of the MSSM. In addition, it allows for
the justification of the charge quantization as long as the matter
lives on the visible brane or in the bulk. Due to $d=6$
proton-decay operators the first and the second family of the
matter fields have to originate from the bulk.
\begin{acknowledgments}
The author would like to thank Stephen M.\ Barr, Hyung Do Kim, and
Stuart Raby for reading this manuscript and for valuable comments.
\end{acknowledgments}

\end{document}